\documentclass{PoS}

\title{Evidence for a pion condensate formation in pp interactions at U-70}

\ShortTitle{Evidence for a pion condensate formation in pp interactions at U-70}

\author{\speaker{Elena Kokoulina}
        \thanks{On behalf of the SVD2 Collaboration.}\\
        Joint Institute for Nuclear Research\\
        E-mail: \email{kokoulin@sunse.jinr.ru}}

\abstract{A high multiplicity study (project Thermalization, experiment E-190) is carried out on U-70 accelerator at IHEP (Protvino, Russia). This project is aimed at searching for collective phenomena. It is known that pions are mainly produced at the 50 GeV-proton beam accelerator. Their mean energy decreases while multiplicity increasing. Bose-Einstein condensate (BEC) can be formed in this system. The theoretical and experimental studies of BEC have been performed since 70es. Within the framework of an ideal pion gas model M.~Gorenstein and V.~Begun have shown that sharp growth of fluctuations of the neutral pion number will be a signal of BEC formation with the increase of the total multiplicity (neutral and charged particle sum). SVD-2 Collaboration (JINR, IHEP and SINP MSU) investigated fluctuations of the neutral pion number in pp interactions at 50 GeV/c incident beam on U-70 versus the total multiplicity and has revealed noticeable growth of the scaled variance with the total multiplicity increase. The growth of these fluctuations reaches more than 7 standard deviations for the scaled variance at the total multiplicity about 30 pions as opposed to the tendency for the simulated events. This growth has been observed both in the registered photons and restored neutral pions.
}

\FullConference{36th International Conference on High Energy Physics\\
        4-11 July 2012\\
        Melbourne, Australia}

\begin{document}

\section{Introduction}
Experiments at LHC have given evidence of similarity of multiple production mechanisms in proton interactions at high multiplicities \cite{CMS} and central collisions of relativistic heavy ions (RHIC). Studies in the high multiplicity (more than the mean value) region are carried out on U-70 accelerator at IHEP (Protvino). The experimental setup Spectrometer with Vertex Detector (SVD-2) used for these investigations consists of a precision silicon vertex detector, a drift tube tracker, a magnetic spectrometer with sixteen proportional chambers and a magnet, a Cherenkov counter and an electromagnetic calorimeter (EMCal). SVD-2 Collaboration (JINR, IHEP and SINP MSU) \cite{SVD} is aimed at searching for collective phenomena in the high multiplicity area. It is known that mainly pions are produced at U-70 energies. Their mean energy decreases with multiplicity increase. Bose-Einstein Condensate (BEC) can be formed in this system.

The theoretical and experimental studies of BEC have been performed since 70es. Within the framework of an ideal pion gas model M.~Gorenstein and V.~Begun \cite{Goren} have shown that sharp growth of fluctuations of the neutral pion number will be a signal of BEC formation with the increase of the total multiplicity (sum of the neutral and charged particle number). They proposed to measure scaled variance, $\omega $, defined as the ratio of the variance of the distribution of the neutral pion number to their mean multiplicity $\omega~=~(<N_0^2>-<N_0>^2)/<N_0>$ at the fixed total multiplicity. In the thermodynamic limit \cite{Goren} this quantity approaches to infinity. It reaches the finite value for the restricted system size formed in the collisions of two protons. Fluctuations of the neutral pion number were investigated in pp interactions on the  50 GeV/c incident beam at U-70 versus the total multiplicity, and the noticeable growth of $\omega $ with the total multiplicity increase, has been revealed \cite{SVD}.

This study is carried out in two stages. At the first stage the topological cross sections are measured. At the second stage the neutral pion number distributions are restored. We could go three orders down on topological cross sections up to $\sim $ 10 nb in comparison with Mirabelle data \cite{Mirab}. The events with high charged multiplicity are extremely rare therefore we have designed a sophisticated trigger to suppress recording of events with the multiplicity smaller than the given value, called a trigger level \cite{Trig}.

The measured topological cross sections are corrected for trigger conditions, detector acceptance, the efficiency of the setup and the reconstruction algorithm. The measurements of charged multiplicity have been fulfilled by using the silicon vertex detector. Comparison of the topological cross sections with the models has shown that the negative binomial distribution (NBD) overestimates the experimental data in the high multiplicity region, $N_{ch}$ > 20, but describes well the region of small multiplicity. A good agreement has been received using the gluon dominance model (GDM) including the fission of gluon sources \cite{GDM}. The neutral pion and photons multiplicities are determined by means of the electromagnetic calorimeter (EMCal).

First indications of the growth of fluctuations of the neutral pion number in the high total multiplicity area were obtained and published one year ago \cite{Fluc}. Thereafter the number of analyzable events was increased by two times. A new more detailed Monte Carlo simulation of the apparatus performance has been carried out and the improvement of the photon reconstruction algorithm has been reached. Our preliminary conclusion remains in force: the growth of a scaled variance begins from total multiplicity $N_{tot}$=18 and gets 7 standard deviations at the total multiplicity about 30 pions as opposed to the tendency for the simulated events.

   \begin{figure}
   \resizebox{0.95\textwidth}{!} {
   \includegraphics{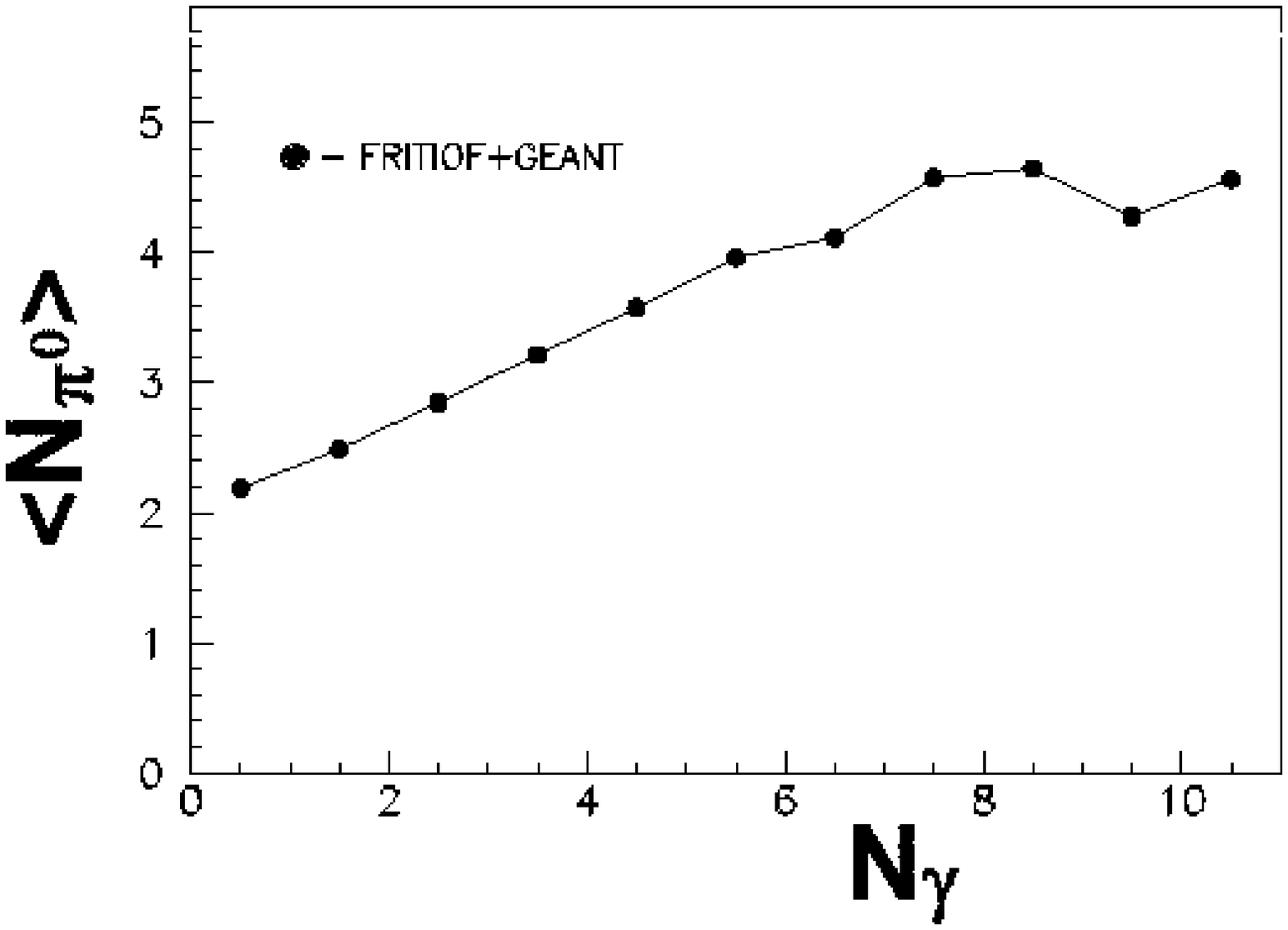}
   \includegraphics{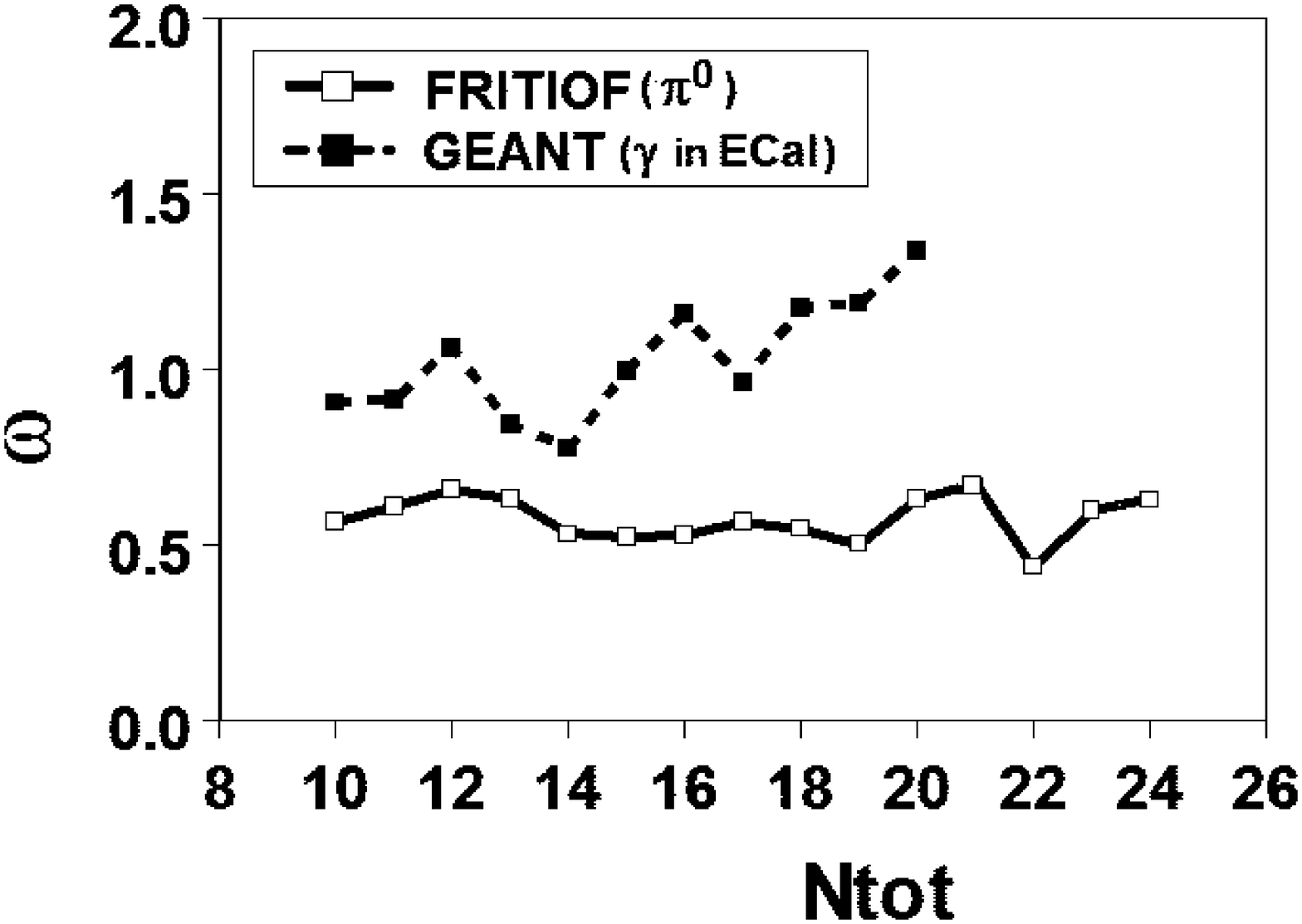}}  
   \caption{Monte Carlo code simulation. [Left] The mean multiplicity of neutral pions
   versus the photon number detected in EMCal.
   [Right] The scaled variance, $\omega $, versus $N_{tot}$
    for photons ($\blacksquare $) and $\pi ^0$-mesons ($\square $).}
   \label{fig:1}
   \end{figure}

\section{Neutral pion restoration}
 Owing to the restricted aperture of EMCal and the threshold energy of gamma-quantum registration, the restoration of $\pi ^0$- mesons in every event is impossible. Their multiplicity is restored by means of Monte Carlo simulation. Now for this purpose we have used Monte Carlo event generator FRITIOF7.02. For 10 mil simulated inelastic events the linear dependence between the mean multiplicity of produced $\pi ^0 $- mesons and the number of the photons registered from their decays is corroborated (the left panel of Figure~1) again. Such dependence allows one to determine the scaled variance of neutral pions indirectly by means of the measured scaled variance of $\gamma $- quantum straight way. The scaled variance for Monte Carlo events is shown in the right panel of Figure~2. The small increasing of $\omega $ versus the total multiplicity is observed for photons ($N_{tot}=N_{ch}+N_{\gamma }$) and  at the same time its value is constant for neutral pions ($N_{tot}=N_{ch}+N_0$) in the investigated area.

Monte Carlo simulation (PYTHIA5.6, $10^7$ events) is used to restore a neutral pion multiplicity. These events are divided in the groups of events on the charged multiplicity, $N_{ch}$. Further we analyze  every group.  These groups consist of subgroups with the defined number of the registered photons (observed in EMCal), $N_{\gamma}$ = i. It is obvious that almost all these photons appear after decays of neutral pions (Monte Carlo simulation supports this conclusion).

Let $N_{ev}(i)$ be the number of all events with i registered photons and $N_{ev}(i,j)$ - the number of the events found in the same subgroup which appeared after the decay of $N_0$ = j neutral pions. We define the matrix coefficients  $c_{ij}=N_{ev}(i,j)/N_{ev}(i)$.  These coefficients  determine the share of events with j neutral pions among the events registered as events with i photons. The simulation allows one to obtain $c_{ij}$ for $N_{\gamma}\leq 10$ and $N_{ch}\leq 14$ only. This is because the high multiplicity events are rare Monte Carlo events. Regularities of probabilities of $c_{ij}$ are used to extrapolate them to $N_{\gamma}>$ 10 and $N_{ch}>$ 14 region and the full sample of $N_{ev}(N_{tot}, N_{ch}, N_0)$ is obtained.

Using these probabilities we can redistribute the number of experimental events observing photon multiplicity into samples on the number of events with different multiplicity of $\pi ^0$ - mesons. This is the inverse task. After that the distributions of the neutral pion number at the total multiplicity can be defined. The average number of neutral pions $<N_0>$ after their restoration (Fig.~2) is in the agreement with Mirabelle data at 70 GeV \cite{Bor} (a small multiplicity region). This agreement justifies that the procedure of this restoration has been carried out correctly. At high charged multiplicity the discrepancy between two Monte Carlo codes, PYTHIA6.2 and FRITIOF7.02 predictions is stipulated by different particle production mechanisms designed in these codes.

In the present analysis about one million events of pp-interactions have been selected, two times more than in the previous one \cite{Fluc}. Photon reconstruction has been improved. It consisted of searching for the cell consisted of 5 $\times $ 5 (in the previous case the 3 $\times $ 3 cell was chosen) signal clusters in EMCal analyzing them with photon criteria. The transverse size of the glass is equal to the Moli\`{e}re radius. Besides, the following procedure is carried out. The experimental events were divided into groups according to charged particle number, $N_{ch}$. Every group is divided into subgroups of events on EMCal registered photon number, $N_{\gamma }$. In each of these subgroup ($N_{\gamma }$ is fixed) the number of events with a certain number of $\pi ^0$ - mesons is restored.

The transition from the number of events with the registered photon multiplicity to the number of the events with the restored neutral pion multiplicity is the inverse task. This task is resolved by using $c_{ij}$ coefficients and the multiplicity distributions of neutral mesons, $N_0$ at different values of $N_{tot} = N_{ch} + N_0$ are found. To analyze neutral pion number distributions at different values of the total multiplicity, $N_{tot}$, the variable $n_0 = N_0 / N_{tot}$, (the region of change [0, 1]) is used. These distributions $r_0(n_0)$ are presented in the left panel of Figure~3 at $N_{tot}$ = 10, 11, ..., 25 + 26 + 27. The data for $N_{tot}$ = 25, 26, and 27 are combined due to small statistics.
   \begin{figure}
       \centerline{
       \includegraphics[width=9. cm,height=6.5 cm]
                                   {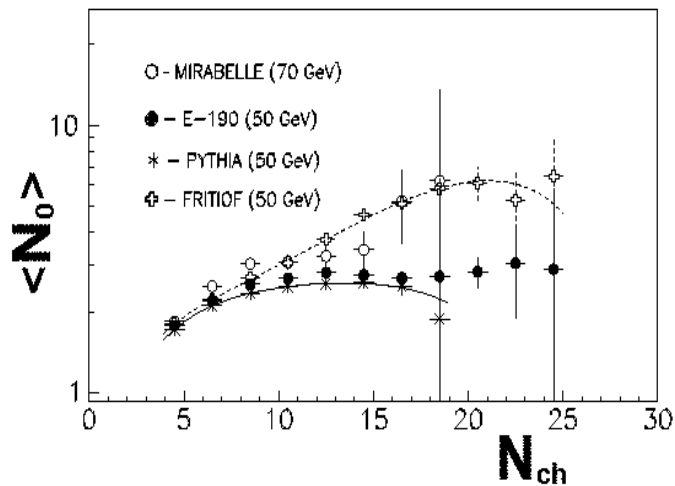}}
   \caption{The dependence of the average number of neutral pions $< N_0 >$ on charge multiplicity for Monte Carlo (PYTHIA, FRITIOF) events, Mirabelle and SVD-2 data.}
   \label{fig:2}
   \end{figure}

   \begin{figure}
   \resizebox{0.95\textwidth}{!} {
   \includegraphics{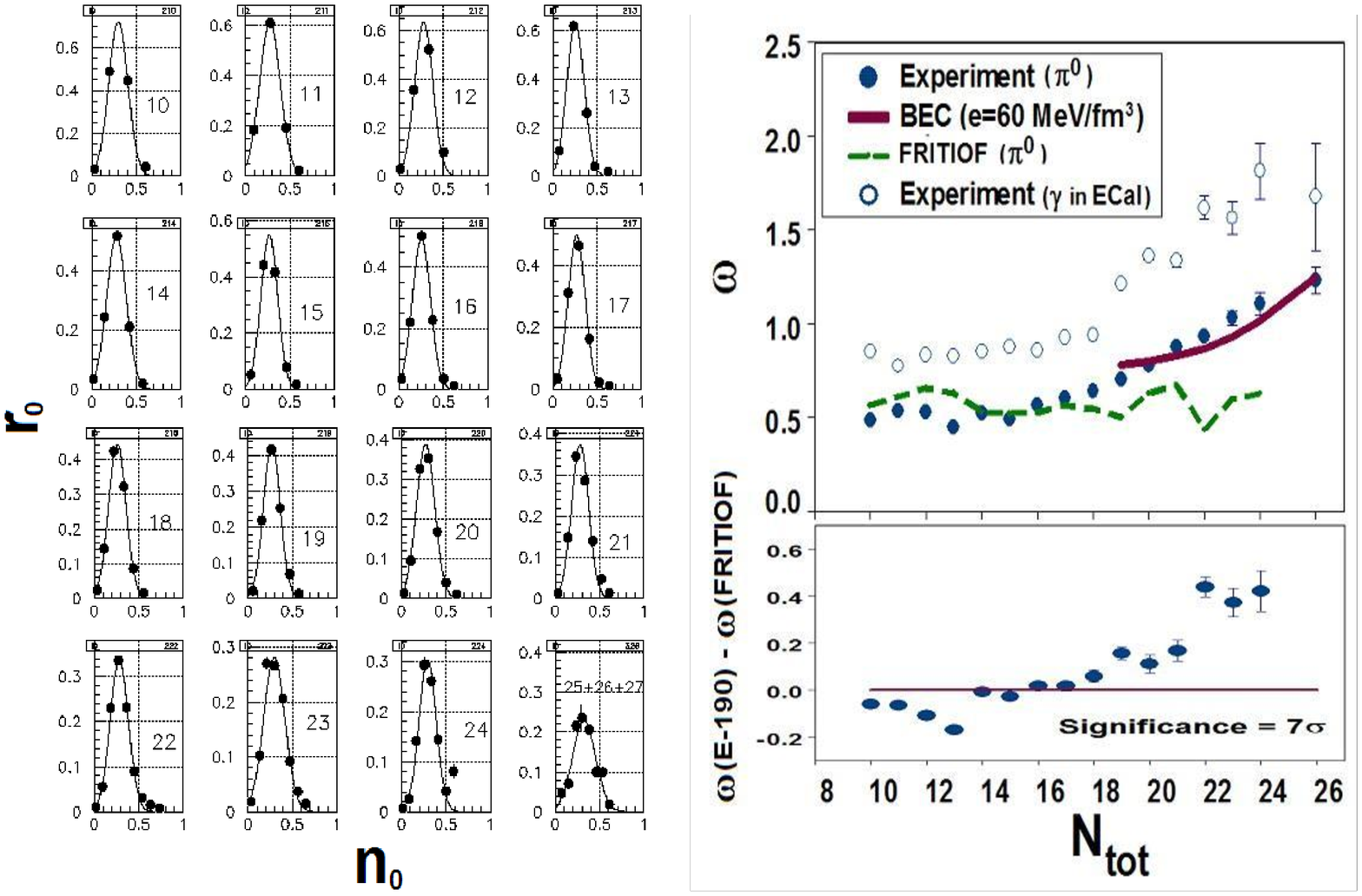}}
   \caption{[Left] The neutral pion number distributions $r_0(n_0)$ versus scaled multiplicity, $n_0$ at different values $N_{tot}$ (10, 11, ..., 25+26+27).
   [Right] (Top) The measured scaled variance $\omega $ versus $N_{tot}$
    for $\pi ^0$-mesons ($\bullet $), photons ($\circ $),
    MC code FRITIOF7.02 (the dashed curve) and theoretical prediction
    (the solid curve) \cite{Goren} for energy density $\epsilon $ =60 MeV/fm$^3$. $N_{tot}=N_{ch}+N_0$ for $\pi^0$-mesons and $N_{tot}=N_{ch}+N_{\gamma }$ for photons. (Bottom) The difference of experimental and simulated $\omega $'s for $\pi^0$-mesons.}
   \label{fig:3}
   \end{figure}

\section{Neutral pion number fluctuations}
The distributions $r_0(n_0)$  in the left panel of Figure~3 are parameterized by the Gauss function to find their mean value and standard deviation. After that the scaled variance is calculated. In the right panel of Figure~3 (top) the experimental, Monte Carlo simulation and model predictions of $\omega $ are presented. The experimental scaled variance agrees well with the magnitude of $\omega $ defined on the simulated events at $N_{tot}$ < 18. At the same time we have revealed the significant growth of it, reaching more than 7 standard deviations at $N_{tot}$ $\simeq$ 30 (the right panel of Figure~3, bottom) as opposed to the tendency for the simulated events. This growth has been observed both in the registered photons ($N_{tot} = N_{ch} + N_{\gamma }$), and restored neutral pions ($N_{tot} = N_{ch} + N_{ch}$). The theoretical predictions done by V.~Begun and M.~Gorenstein \cite{Goren} under different conditions (the size of the  system, the energy density, the pion number density and others) in the high total pion multiplicity area confirm the Bose-Einstein condensate formation since we have observed the growth of the scaled variance.

 The critical point of pion condensation is determined in statistical physics: $E_{crit}=3.3 (h^2/m_{\pi }) \rho ^{2/3} $
 \cite{Land}. The density $\rho $ is equal to 0.2 $fm^{-3}$ if the interaction region size of two protons $\sim $ 3 fm and $N_{tot}$ = 36 (the maximal observable number of pions). In this case the critical energy is equal to $E_{crit}$ $\simeq $ 100 MeV. At the 50-GeV proton beam and $N_{tot}$ = 30, the mean energy of the pion,
 $$E_{\pi } = \left (E_{cms} - 2 m_{n} - N_{tot} m_{\pi }\right)/N_{\pi },$$
 is equal to 120 MeV ($m_n$ -- the nucleon mass, $N_{\pi }$ -- the number of pions and $m_{\pi }$ -- their mass). This value is compatible with $E_{crit}$. Thus the experimental observable growth of scaled variance at U-70 for the registered $\gamma $-quanta and restored neutral pion multiplicity can indicate BEC formation in the pion system at high multiplicity events.

 At present we are planning to study the soft photon ($E_\gamma $ < 100 MeV) yield versus neutral, charged and total multiplicities. The experiments carried out during last decades have pointed out to the excess to its yield in the comparison with the theoretical estimations. The exhaustive explanation of this collective phenomenon is absent. In the approach developed by S. Barshay \cite{Barsh} the excess of the soft photon yield can be related with BEC formation in the pion system. We will also increase statistics to move forward to a much higher multiplicity region.

I appreciate all participants of SVD-2 Collaboration for active and
fruitful work and thank the ICHEPP 2012 Organizing Committee for the exciting friendly atmosphere and good conditions of work and the communication during all the stay in Melbourne.


\begin{thebibliography}{99}
   \bibitem{CMS}
   CMS Collaboration, \emph{Observation of Long-Range Near-Side Angular Correlations in
   Proton-Proton Collisions at the LHC}, JHEP 09 (2010) 091 [hep-ex/1009.4122].
   \bibitem{SVD}
   E.~N.~Ardashev et al. SVD-2 Collaboration. (in Russian)
   \emph{Proton interactions with high multiplicity}. IHEP Preprint 2011-4
   (2011) [hep-ex/1104.0101].
   http://web.ihep.su/library/pubs/all-w.htm.;
   E.~N.~Ardashev et al. SVD-2 Collaboration. (in Russian)
   \emph{Neutral-Pion Fluctuations at High Multiplicity in pp Interactions
   at 50 GeV}.
   IHEP Preprint 2011-5 (2011). [hep-ex/1104.3673]
    http://web.ihep.su/library/pubs/prep2011/11-5-w.htm.
   \bibitem{Goren}
   V.~V.~Begun and M.~I.~Gorenstein, \emph{Bose-Einstein condensation of pions in high multiplicity events},
   Phys.~Lett. \textbf{B} 653 (2007) 190;
   V.~V.~Begun and M.~I.~Gorenstein, \emph{Bose-Einstein Condensation in the Relativistic Pion Gas:
   Thermodynamic Limit and Finite Size Effects}.
   Phys.~Rev. \textbf{C}77 (2008) 064903.
   \bibitem{Mirab}
   V.~V.~Ammosov et al.,
   \emph{Average charged particle multiplicity and topological
   cross-sections in 50-GeV/c and 69-GeV/c p p interactions}.
   Phys. Lett. \textbf{B} 42 (1972) 519.
   \bibitem{Trig}
   A.~Avdeichikov {\it et al.},  	
   \emph{A trigger of events with a high multiplicity of charged particles at the SVD-2 setup}.
   Instrum.~Exp.~Tech. \textbf{2} (2011) 15.
   \bibitem{GDM}
   E.~S.~Kokoulina. Acta~Phys.~Polon.  	
   \emph{Description of pp interactions with very high multiplicity at 70-GeV/c.}
   \textbf{B} 35 (2004) 295; \emph{Gluon dominance model}. AIP Conf. Proc. \textbf{828} (2006) 81;
   E.~A.~Kuraev, S.~Bakmaev, E.~S.~Kokoulina, \emph{Azimuthal correlation of gluon jets created in proton-antiproton annihilation.}
   Nucl.~Phys. \textbf{B} 851 (2011) 551.
   \bibitem{Fluc}
   E.~S.~Kokoulina (On behalf of the SVD-2 Collaboration),
   \emph{Neutral Pion Fluctuations in pp Collisions at 50 GeV by SVD-2}. Progr. Theor. Phys., {\bfseries 193} (2012) 306;
   V.~N.~Ryadovikov (On behalf of the SVD-2 Collaboration),
   \emph{Neutral-Pion Fluctuations at High Multiplicity in pp Interactions
   at 50 GeV}. Phys.~Atom.~Nucl., {\bfseries 75} (2012) 989.
   \bibitem{Bor}
   M.~Boratov et al.
   \emph{ Gamma Production and Multiplicity Correlations Between Neutral and Charged Pions in p p Interactions at 69-GeV/c}.
   Mirabelle Collaboration. Nucl.~Phys. \textbf{111} (1976) 529.
   \bibitem{Land}
   L.~D.~Landau and I.~M.~Lifshitz. Vol. 5. Statistical physics, part 1 (3ed., Pergamon, 1980).
   \bibitem{Barsh}
   Saul~Barshay, \emph{ANOMALOUS SOFT PHOTONS FROM A COHERENT HADRONIC PHASE
   IN HIGH-ENERGY COLLISIONS}. Phys.~Lett. \textbf{B} 227 (1989) 279.
\end{thebibliography}
\end{document}